\documentclass[aps,twocolumn,pop]{revtex4}
\usepackage{graphicx}
\usepackage{amsmath}
\usepackage{amssymb}
\usepackage{amscd}
\usepackage{bbm}
\usepackage{epstopdf}
\usepackage{times,mathptmx}

\newcommand{\comment}[1]{}
\newcommand\Ai{\mathrm{Ai}}

\begin{document}

\title{The lower hybrid wave cutoff: a case study in eikonal methods}


\author{A.~S.~Richardson}
\affiliation{Plasma Science and Fusion Center, MIT, Cambridge, Massachusetts 02139, USA}

\author{P.~T.~Bonoli}\affiliation{Plasma Science and Fusion Center, MIT, Cambridge, Massachusetts 02139, USA}
\author{J.~C.~Wright}\affiliation{Plasma Science and Fusion Center, MIT, Cambridge, Massachusetts 02139, USA}

\begin{abstract}
Eikonal, or ray tracing, methods are commonly used to estimate the propagation of radio frequency fields in plasmas.  While the information gained from the rays is quite useful, an approximate solution for the fields would also be valuable, e.g., for comparison to full wave simulations.  Such approximations are often difficult to perform numerically, because of the special care which must be taken to correctly reconstruct the fields near reflection and focusing caustics.  In this paper, we compare the standard eikonal method for approximating fields to a method based on the dynamics of wave packets.  We compare the approximations resulting from these two methods to the analytical solution for a lower hybrid wave reflecting from a cutoff.  The algorithm based on wave packets has the advantage that it can correctly deal with caustics, without any special treatment.
\end{abstract}

\maketitle

\section{Introduction}

Eikonal methods---also known as ray tracing or Wentzel--Kramers--Brillouin (WKB) methods---are often used to model the propagation and absorption of radio frequency waves in plasmas (for ray tracing applied to lower hybrid waves see \cite{wersinger:2263,1978PZhTF...4R.800B,0032-1028-24-10-001,bonoli_1984_LH,citeulike:369883}).  The results obtained from ray tracing can be compared to full wave simulations, in order to reproduce, supplement, and better understand those simulations \cite{aps08_paper_jcw}.  Since the ray tracing method was devised as an approximate method for problems where the wavelength is short compared to gradients in the medium, there are problems which cannot be accurately solved using ray tracing.  There are also cases where diffractive effects like focusing and reflections from cutoffs are not handled well by traditional ray tracing methods.  These caustics are often interpreted as a breakdown of the eikonal approximation used to justify ray tracing, and therefore an indication that the ray tracing results are not valid \cite{cardinali:112506}.  However, more advanced eikonal methods have been developed which allow ray tracing to be used even in the presence of caustics.  These methods, based primarily on the work of Maslov \cite{maslov:book}, are routinely and successfully applied to wave problems in various fields, including atomic, molecular, and optical (AMO) physics \cite{knudson:5703,citeulike:745805} and geophysics \cite{Chapman:2002ff,thomson_chapman:intro}. Because these advanced eikonal methods are able to deal with caustics, it becomes possible to  construct approximate wave fields based on ray tracing data.  

However, the techniques based on the work of Maslov require careful application, and special attention must be paid to the regions around caustics.  Different types of caustics require various methods to deal with them, and it becomes a ``labor of love''\cite{Delos:quote} to perform the field reconstruction.  Because of the difficulty of properly applying the Maslov methods, it is quite challenging to create an numerical algorithm for field reconstruction based on these techniques.  

In this paper we have, for the first time that we are aware of, employed these methods to reconstruct RF wave fields in situations relevant to realistic fusion plasmas.  The problem of a lower hybrid wave reflecting from the density cutoff is considered.  Given certain approximations, this problem can be solved analytically for a cold plasma slab model.  We first describe the model system and the analytical solutions, which can be compared to the approximate solutions that will be presented.  We then use the standard Maslov technique to construct an eikonal solution which is valid even in the region near the caustic.  This will illustrate the difficulties that arise when trying to use standard eikonal methods to automatically construct approximate solutions numerically.  We then describe a technique for constructing fields that is based on an approximation for wave packet dynamics.  Application of this method to the lower hybrid cutoff problem shows how it is well suited for use as a numerical algorithm for the construction of approximate solutions, even in the presence of caustics.  The paper ends with a comparison of the two approximate solutions to the analytical solution, and a discussion of the results.

\section{The lower hybrid cutoff\label{sec:exact}}

In this section, we will describe the problem of a lower hybrid wave reflecting from the density cutoff, and derive the analytical solution.  In Sec.~\ref{sec:compare}, we will use this solution as a benchmark with which to compare our eikonal solutions.

As in Stix \cite{stix:waves}, the equation we want to solve is
\begin{align}\label{eq:prob}
-\nabla \times (\nabla \times {\bf E}) + \frac{\omega^2}{c^2} {\bf K} \cdot {\bf E} = 0,
\end{align}
which is the wave equation for the electric field ${\bf E}$ in a plasma with dielectric tensor ${\bf K}$.  The time dependence in this equation has already been removed through the substitution $i\partial_t \rightarrow \omega$.  This wave equation can be written in a compact form as
\begin{align}\label{eq:wave}
\hat {\bf D}({\bf x},\nabla) \cdot {\bf E}({\bf x}) = 0,
\end{align}
where $\hat {\bf D}$ is the wave operator, which is a function of the noncommuting operators ${\bf x}$ and $\nabla$.  The ambiguity in the ordering of these operators is resolved by using the Wigner-Weyl formalism, a description of which can be found in Appendix B of Ref.~\cite{citeulike:703463}.  This formalism provides a well defined process for computing the dispersion matrix ${\bf D}({\bf x},{\bf k})$ associated with an arbitrary wave operator $\hat {\bf D}$.  However, for the problem to be considered here, this process reduces to making the substitution
$-i \nabla \rightarrow {\bf k} = {\bf N} \omega/ c$,
since the ordering of the multiplication and gradient operators is unambiguous for this model.  We will now give the detailed description of our model and its dispersion matrix.

\subsection{Description of the model}

Consider a two-dimensional (2D) cold plasma slab model with a linear density gradient $n(x,z) = n' x$.  The magnetic field is uniform, and points in the $z$ direction: ${\bf B}(x,z) = B_0 \hat z$.  In this paper, we will use the parameters $B_0 = 5.5\,$T, and $n' = 3\times 10^{17} {\rm m}^{-4}$, $f = 4.6\,$GHz.   We will use the cold plasma dispersion matrix from Stix \cite{stix:waves}:
\begin{align}\label{eq:stix}
{\bf D} = 
\left[
\begin{array}{ccc}
S +N_x^2 - N^2 & -iD + N_x N_y & N_x N_{z}  \\
 iD+ N_x N_y  & S + N_y^2 -N^2 &  N_y N_z \\
 N_x N_{z} &  N_y N_z  &  P + N_z^2 -N^2
\end{array}
\right],
\end{align}
where 
\begin{align}
S &= 1-\sum_j \frac{\omega_{pj}^2}{\omega^2-\Omega_j^2} \\
D &= \sum_j  \frac{\epsilon_j \Omega_j}{\omega}\frac{\omega_{pj}^2}{\omega^2-\Omega_j^2} \\
P &= 1- \sum_j \frac{\omega_{pj}^2}{\omega^2},
\end{align}
and $\omega_{pj}$ and $\Omega_j$ are, respectively, the plasma and cyclotron frequencies of the $j^{th}$ particle species.  Also, $\epsilon_j$ is the sign of the charge of the $j^{th}$ species.  The frequencies are given by
\begin{align}
\omega_{pj}^2 = \frac{n e Z_j^2}{\epsilon_0 m_j} \,\,\,\,{\rm and } \,\,\,\,
\Omega_j = \left\vert \frac{Z_j e {\bf B}}{m_j}  \right\vert .
\end{align}
Since in this model the magnetic field is assumed to be constant in space while the density is allowed to vary, all of the spatial variation in {\bf D} is contained in the plasma frequencies $\omega_{pj}$.

In this paper, we are interested in the reflection of a wave from the slow wave cutoff at $P(x)=0$.  In this region, the spatial variation which we are most interested in is the variation in $P(x)$.  Here, $S(x) \simeq 1$ and $D(x) \simeq 0$.  In order to be able to derive an analytical solution to our problem, we will modify the cold plasma dispersion relation by setting $D=0$ and $S=1$.  While this does change the problem, it retains the physics of interest [the variation in $P(x)$ which gives the slow wave cutoff], and it allows us to derive an exact solution that we can compare to our numerical field reconstructions given in the later sections.  These modifications ($D=0$ and $S=1$) will be retained throughout the remainder of the paper.

\begin{figure}[tb]
\includegraphics[width=8.5cm]{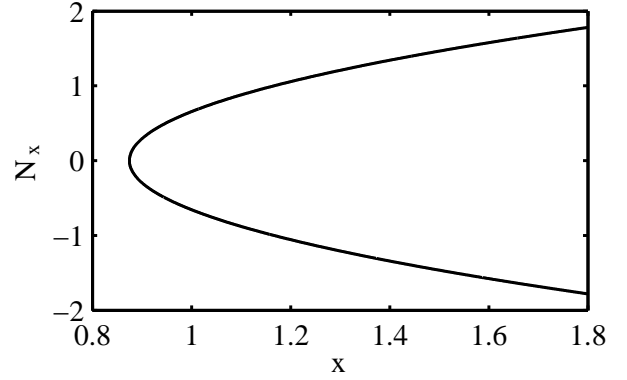}
\caption{Dispersion function for the slab model used in this paper; $N_y = 0$, $N_z = 2$.}
\label{fig:disp}
\end{figure}

With these modifications, the dispersion function for our model becomes:
\begin{align}
\mathcal{D}(x,z,N_x,N_z) &= \det[{\bf D}(x,z,N_x,N_z)] \\
&\!\!\!\!\!\!\!\!\!\!\!\!\!\!\!\!\!\!\!\!\!\!\!\!\!\!\!\!
=(1-N^2) \left[(1-N_z^2)P(x) - N_x^2 - N_y^2\right],
\end{align}
where
\begin{align}
N^2 = N_x^2 + N_y^2 + N_z^2.
\end{align}
Solving for $\mathcal{D}(x,z,N_x,N_z) = 0$ gives the dispersion relation for this problem.  Figure \ref{fig:disp} shows the dispersion function for the case $N_y = 0$, $N_z = 2$. 

\subsection{The analytical solution}

In this paper we studying the reflection of a lower hybrid wave from the slow wave cutoff.  For such a reflection, the dominant spatial variation is in the direction perpendicular to the cutoff (the $x$ direction), so we have modeled our plasma as being uniform in the $z$ direction.  Since there is no variation in the $z$ direction, we can Fourier transform Eqn.~(\ref{eq:prob}) in $z$, giving us a set of equations for ${\bf E}(x;N_z)$.  These equations can also be obtained from the dispersion matrix in Eqn.~(\ref{eq:stix}) by making the substitution $N_x \rightarrow \hat N_x = -i(c/\omega) \partial_x$, which gives
\begin{align}
\left[
\begin{array}{ccc}
1 -N_y^2 - N_z^2 &  \frac{-icN_y}{\omega} \partial_x &\frac{-icN_z}{\omega} \partial_x  \\
\frac{-icN_y}{\omega} \partial_x  & 1 -N_z^2 +\frac{c^2}{\omega^2}\partial_x^2&  N_y N_z \\
\frac{-icN_z}{\omega}\partial_x &  N_y N_z  &  P(x) + N_y^2 +\frac{c^2}{\omega^2}\partial_x^2
\end{array}
\right]\notag
\\
\cdot
\left[
\begin{array}{c}
E_x(x;N_z) \\
E_y(x;N_z) \\
E_z(x;N_z)
\end{array}
\right] = 0.
\end{align}
Once these equations are solved for each $N_z$, we can form a general solution to the problem by summing each $N_z$ mode with some arbitrary spectral density $F(N_z)$:
\begin{align}\label{eq:fourier_sol}
{\bf E}(x,z) = \int  F(N_z) e^{i\frac{\omega}{c} N_z z} {\bf E}(x; N_z) \, dN_z.
\end{align}
We can now solve for each Fourier mode independently, and we obtain a solution which can be written in terms of the Airy function $\Ai$ and its derivative $\Ai'$.  The vector components of the solution are:
\begin{align}\label{eq:mode_sol}
E_x(x;N_z) &= -\frac{ic}{\omega} \alpha N_z (\partial_x P) \Ai'(\tilde x), \notag\\
E_y(x;N_z) &= \frac{N_y N_z}{N_z^2 - 1} \Ai(\tilde x), \\
E_z(x;N_z) &= \Ai(\tilde x),\notag
\end{align}
where
\begin{align}
\tilde x &= -\alpha \left[ (1-N_z^2) P(x) -N_y^2\right] ,
\end{align}
and
\begin{align}
\alpha^3 &= \left(\frac{\omega}{c (\partial_x P) (1-N_z^2)}\right)^2.
\end{align}
Note that for our linear density profile, $(\partial_x P)$ is a negative constant, and therefore $\alpha$ is a constant.  Also, since we will take $|N_z|>1$, we have that $\alpha$ is a positive real number.  In order to create a beam-like solution, we choose $F(N_z)$ to be a gaussian, with its  peak at $N^{(0)}_z$, and a width of $\sigma_{N_z}$:
\begin{align}
F(N_z) = \exp\left(-\frac{\left(N_z-N^{(0)}_z\right)^2}{2\sigma_{N_z}^2}\right).
\end{align}
\begin{figure}[tb]
\includegraphics[width=8.5cm]{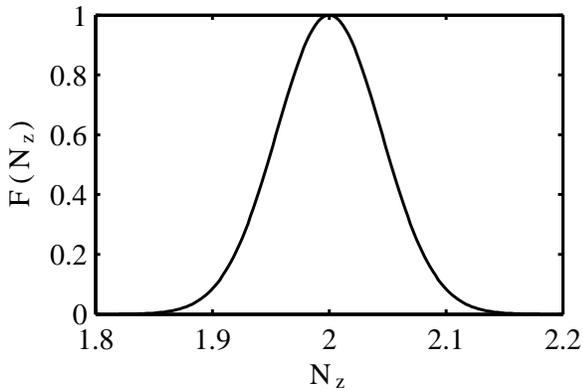}
\caption{Spectrum in $N_z$ for the analytical solution.}
\label{fig:sol_spec}
\end{figure}

\begin{figure}[tb]
\includegraphics[width=8.5cm]{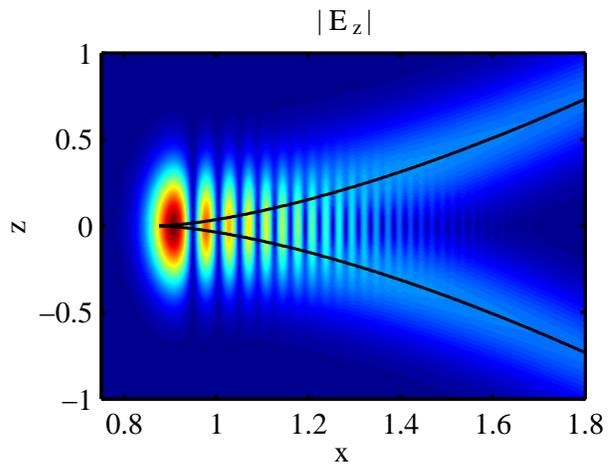}
\caption{(Color online) Absolute value of $E_z$ for the analytical solution.  The black line is the line of stationary phase given in Eqn.~(\ref{eq:st_ph_ray}).}
\label{fig:exact}
\end{figure}

We can get an idea of what the solution with this spectral density will look like by examining the integral in Eqn.~(\ref{eq:fourier_sol}).  Since the spectrum is peaked (see Fig.~\ref{fig:sol_spec}), we can develop some intuition about the solution by applying the stationary phase approximation to the integral.  Rather than using the stationary phase method to obtain an analytical approximation to the solution, we will simply use it to find the curve in the $(x,z)$-plane where we expect the maximum field amplitude.  This curve will describe the center of the resulting ``beam.''  Using the asymptotic expressions for the Airy functions \cite{AS.10}, we obtain the equation for the curve:
\begin{align}\label{eq:st_ph_ray}
z = \pm \frac{2 N_z^{(0)} [-P(x)]^{3/2}}{3 (\partial_x P)\sqrt{{\left(N_z^{(0)}\right)}^2-1}}.
\end{align}
This curve, along with the amplitude of the $E_z$ component of the solution, are shown in Fig.~\ref{fig:exact}.

\section{Standard eikonal field reconstruction\label{sec:eikonal}}

In this section, we apply the standard eikonal approximation to the problem of the reflection of a lower hybrid wave from the cutoff.  We use the method detailed by Maslov and Fedoriuk in Ref.~\cite{maslov:book}.  Rather than giving detailed derivations of the equations used for this approximation, we will simply give the relevant equations and explain how they are used.  For more details, the interested reader is encouraged to refer to Ref.~\cite{maslov:book} or  Ref.~\cite{knudson:5703}, where this method is applied to a quantum scattering problem.

In the standard eikonal method, a family of rays is used to construct a field, which is an approximate solution to the wave equation being studied.  This approximate solution has the eikonal form
\begin{align}
{\bf E}({\bf x},t) = A({\bf x}) {\bf e}({\bf x}) e^{i \Theta({\bf x})-i\omega t}.
\end{align}
The family of rays traces out a surface in $({\bf x},{\bf k})$ phase space, called a Lagrange manifold.  The family of parameterized rays forms a coordinate system on this surface, with the ray ``time'' $\tau$ as one coordinate direction, and the ray label $\beta$ forming the other coordinate.  The individual rays are found by solving Hamilton's equations for the ray position
\begin{align}
\frac{d {\bf x}}{d\tau} &= \frac{d}{d{\bf k}} H({\bf x},{\bf k}), \\
\frac{d {\bf k}}{d\tau} &= -\frac{d}{d{\bf x}} H({\bf x},{\bf k}),
\end{align}
where $H({\bf x},{\bf k})$ is a zero eigenvalue of the dispersion matrix ${\bf D}({\bf x},{\bf k})$, with eigenvector ${\bf e}({\bf x},{\bf k})$.

The spatial gradient of the phase is given by
$\nabla \Theta ({\bf x}) = {\bf k},$
where ${\bf x}$ and ${\bf k}$ are evaluated along the ray.  Given an initial phase $\Theta({\bf x}_0) = \Theta_0$ defined by boundary conditions at ${\bf x}_0$ we can integrate this equation to find $\Theta({\bf x})$:
\begin{align}
\Theta({\bf x}) = \Theta_0 +\int^{\bf x}_{{\bf x}_0} {\bf k}({\bf x}') \, d{\bf x}'.  
\label{eq:phase_int}
\end{align}
Here, ${\bf k}({\bf x})$ refers to the function found by inverting the ray data $\left( {\bf x}(\tau,\beta), {\bf k}(\tau,\beta) \right)$ to obtain $\bf k$ as a function of $\bf x$ along the ray.  In practice, the ray parameter $\tau$ and label $\beta$ are often used for integrating the equation for the phase, since that is simpler than solving for ${\bf k}({\bf x})$.  In this case, we have
\begin{align}\label{eq:phase_int2}
\Theta({\bf x}) = \Theta_0 + \int_{\tau_i}^{\tau_f} \left({\bf k} \cdot \frac{\partial{\bf x}}{\partial\tau}\right) d\tau 
+\int_{\beta_i}^{\beta_f}  \left({\bf k} \cdot \frac{\partial{\bf x}}{\partial\beta}\right) d\beta.  
\end{align}
It turns out that---for single-valued ${\bf k}({\bf x})$---the result of the integral in Eqn.~(\ref{eq:phase_int}) is independent of the path of integration.  Because of this, the limits on the integrals in Eqn.~(\ref{eq:phase_int2}) need only be chosen so that  ${\bf x}(\tau_i, \beta_i) = {\bf x}_0$ and ${\bf x}(\tau_f, \beta_f)$ is the point where $\Theta({\bf x})$ is to be evaluated.  

This integration must be treated slightly more carefully when ${\bf k}({\bf x})$ is double (or multiply) valued.  This situation occurs when Lagrange manifold folds back over itself, which can happen, for example, when rays reflect from a boundary or cutoff (see Fig.~\ref{fig:1d_ray}).  In this case, each sheet of the Lagrange manifold will contribute to the phase of the eikonal solution.  Each sheet can be associated with an eikonal solution; in the case of the reflection from the cutoff, one sheet is associated with the incoming wave, and the other with the reflected wave.  In this case, the integral for $\Theta({\bf x})$ remains the same, and an additional phase shift is added to the outgoing wave.  The value of this phase shift depends on the geometry of the Lagrange manifold through a quantity called the Keller-Maslov index, which will be discussed in more details below.

The formula for the amplitude $A({\bf x})$ of the approximate solution is then written in terms of the determinant of the Jacobian matrix ${\bf J}({\bf x})$ which arises when transforming from the coordinates $(\tau, \beta)$ to the physical coordinates ${\bf x}$:
\begin{align}
A({\bf x}) = \sqrt{\frac{|\det {\bf J}({\bf x}_0)|}{|\det {\bf J}({\bf x})|}} \label{eq:amp_x}
\end{align}
where
\begin{align}
{\bf J}({\bf x}) = 
\left(
\begin{array}{cc}
\frac{\partial x}{\partial\tau}   & \frac{\partial z}{\partial\tau}  \\
\frac{\partial x}{\partial\beta}   & \frac{\partial z}{\partial\beta}  
\end{array}
\right).
\end{align}
When the coordinate transformation becomes singular, the associated field amplitude goes to infinity.  This breakdown in the approximation for the amplitude is due to a singularity in projecting the Lagrange manifold from phase space to $x$-space.  It is often the case, however, that this singularity can be ``repaired'' by finding a local solution by using the eikonal approximation in $k$-space.  This eikonal solution is then Fourier transformed back to $x$-space, giving a local solution.  The incoming and outgoing eikonal waves can then be matched to this local solution in order to obtain an approximate solution without the amplitude singularity at the caustic.  We will first illustrate how this procedure works by performing it for a single $N_z$ mode, which reduces to a one-dimensional (1D) problem.  We will then give the results for a full 2D problem, with a lower hybrid beam reflecting from the cutoff.

\subsection{The eikonal solution: $x$-space}

Consider a single $N_z$ mode, ${\bf E}(x;N_z)$.  The eikonal solution for this mode is straightforward, since it is a 1D problem.  The solution in $x$-space involves three steps: 
\begin{enumerate}
\item Trace the ray, which gives $(x(\tau),k_x(\tau))$
\item Compute eikonal quantities ${\bf e}(\tau)$, $\Theta(\tau)$, and $A(\tau)$
\item Interpolate these quantities onto a grid in $x$ to form the solution
\end{enumerate}
This process will give two eikonal waves, one which approaches the caustic and one travels away from the caustic.  These two waves will be matched to the local solution to give the complete eikonal solution.  The two waves are identified in the algorithm by computing the sign of ${dx}/{d\tau}$ along the ray; it is negative for the incoming part of the ray, and positive for the outgoing part of the ray.

\begin{figure}[tb]
\includegraphics[width=8.5cm]{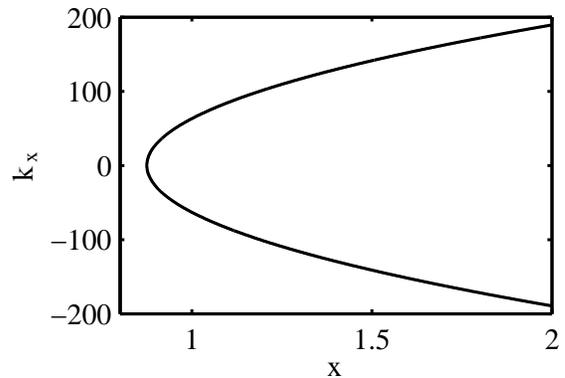}
\caption{The ray $(x(\tau),k_x(\tau))$ for the 1D eikonal solution.}
\label{fig:1d_ray}
\end{figure}

\begin{figure}[tb]
\includegraphics[width=8.5cm]{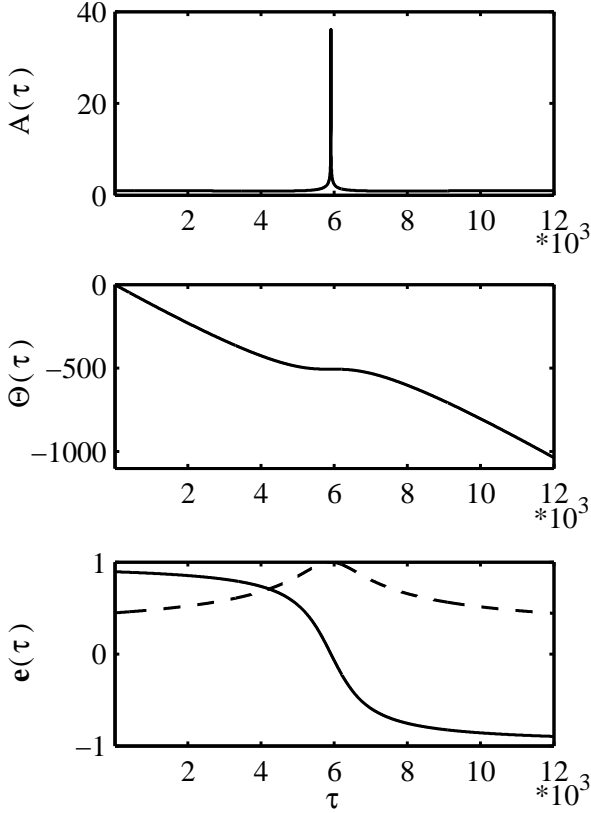}
\caption{The eikonal amplitude, phase, and polarization vector computed along the ray.  Notice that the amplitude blows up at the caustic near $\tau=6000$.  In the plot of the polarization, the solid line is $e_x(\tau)$, and the dashed line is $e_z(\tau)$.}
\label{fig:1d_amp}
\end{figure}

While this model is simple enough that explicit formulas can be written for the eikonal solution, we computed the solution numerically since the point of this calculation is to compare numerical algorithms.  In Fig.~\ref{fig:1d_ray} the computed ray is shown.  Fig.~\ref{fig:1d_amp} shows the eikonal amplitude, phase, and polarization as computed along the ray.  Notice that the amplitude blows up at the caustic near $\tau = 6000$.

\subsection{The local solution: eikonal in $k_x$-space}

The eikonal solution in $x$-space diverges at the caustic, but this divergence can be ``repaired'' by estimating the solution in $k_x$-space.  Here, the eikonal approximation can again be made:
\begin{align}
{\bf E}(k_x,t) = A(k_x) {\bf e}( k_x) e^{i\Theta(k_x) - i\omega t}.
\end{align}
This approximate solution is valid even when the ray is near the caustic point at $k_x =0$, since projecting from phase space to $k_x$-space is well defined.  The projection singularity encountered in $x$-space is not a problem in this representation.

First, consider the amplitude and polarization of this approximation.  These quantities can be evaluated along the ray as functions of $\tau$, and then interpolated onto $k_x$ to get the overall amplitude and polarization as functions of $k_x$.  The equation for the amplitude as a function of $k_x$ takes the same form as Eqn.~(\ref{eq:amp_x}), with $k_x$ taking the place of $x$.  This gives
\begin{align}
A(k_x) {\bf e}(k_x) = A_0 \left\vert \frac{dk_x}{d\tau}\right\vert^{-1/2} {\bf e}(k_x(\tau),x(\tau)).
\end{align}
For the case $N_y=0$ that we are considering, there are two components of ${\bf e}$ which are nonzero.  This means that generically, we need to construct two eikonal functions, one for each nonzero component of the polarization.  However, we can use the wave equation [Eqn.~(\ref{eq:wave})] and the dispersion matrix in Eqn.~(\ref{eq:stix}) to write a simple relationship between $E_x(x)$ and $E_z(x)$:
\begin{align}\label{eq:Ex}
E_x(x) = \frac{ic N_z}{\omega (1-N_z^2)} \, \partial_x E_z(x).
\end{align}
So, we only need to construct the eikonal approximation for $E_z(x)$, and then take a derivative to obtain the approximation for $E_x(x)$.  Therefore, when considering the amplitude and polarization $A(k_x) {\bf e}(k_x)$, we are primarily interested only in the $z$ component, $A(k_x) e_z(k_x)$.  Analytically, we expect this product to be constant, and when it is computed from the ray tracing data, it is constant to within a small numerical error.

We now have
\begin{align}
E_z(k_x) = \tilde A_0 e^{i \Theta(k_x)},
\end{align}
where $\tilde A_0$ is a complex, constant amplitude, and
\begin{align}
\Theta(k_x) = \int^{k_x} x(k_x')\, dk_x'.
\end{align}
As can be seen by the shape of the ray in Fig.~\ref{fig:1d_ray}, we can make a good approximation to the curve $x(k_x)$ by assuming it is quadratic:
\begin{align}
x(k_x) \approx x_0 + \gamma k_x^2,
\end{align}
where the constants $x_0$ and $\gamma$ are computed numerically, and found to be
\begin{eqnarray}
x_0 =  0.8747 \quad {\rm and } \quad \gamma = 3.137\times 10^{-5}.
\end{eqnarray}
Some care must be taken with the sign when integrating to get $\Theta(k_x)$, since the starting point of the integral is the starting point of the ray, which is a large positive value of $k_x$.  This gives an overall minus sign when doing the integral.  Putting all of this together, we get, up to an overall complex constant amplitude, 
\begin{align}
E_z(k_x) = \exp\!\left[i \left(-x_0 k_x - \tfrac{1}{3} \gamma k_x^3 \right)\right].
\end{align}

Now, to get the solution in $x$ space, take the Fourier transform of this function.  The result can be written in terms of the Airy function, giving the local solution
\begin{align}\label{eq:Ez_local}
E_z(x) =  A_0 \, \Ai\! \left(-\gamma^{-1/3} (x - x_0)\right),
\end{align}
where $A_0$ is a complex normalization that will be set by matching to the incoming wave.  We can now use Eqn.~(\ref{eq:Ex}) to find the $x$ component of the local solution, and we get
\begin{align}\label{eq:Ex_local}
E_x(x) = -\frac{i A_0 c N_z \gamma^{-1/3}}{\omega (1-N_z^2)} \,\, \Ai'\!\left(-\gamma^{-1/3} (x - x_0)\right).
\end{align}

\subsection{The matched solution}

We now have numerical solutions for the incoming and outgoing eikonal waves, which we can match to the local solutions in Eqns.~(\ref{eq:Ez_local}) and (\ref{eq:Ex_local}) to obtain the complete eikonal solution.  In order to obtain the correct matching between the local and eikonal solutions, we will need to calculate the phase shift between the incoming and outgoing waves.  This phase shift for our 1D problem is given by the equation
\begin{align}
\phi = -\frac{\pi \mu}{2}
\end{align}
where $\mu$ is the Keller-Maslov index
\begin{eqnarray}
\mu = {\rm inerdex}\left(\frac{\partial x}{\partial k_x}\right)_{\rm in}
- {\rm inerdex}\left(\frac{\partial x}{\partial k_x}\right)_{\rm out},
\end{eqnarray}
and ${\rm inerdex}({\bf M})$ is the number of negative eigenvalues of the matrix $\bf M$ (for a more complete discussion of the Keller-Maslov index in the context of constructing approximate fields, see \cite{knudson:5703}).  For our problem, the slope of $x(k_x)$ is positive for the incoming wave and negative for the outgoing wave, so we have $\mu = 0-1 = -1$.

It now remains to combine the incoming and outgoing waves in such a way as to form the complete eikonal solution.  This can be achieved by combining them with a switching function, so that the local solution is used in the region of the caustic, and the eikonal solutions are used away from the caustic.  The code developed for this paper uses a hyperbolic tangent function to achieve this switching:
\begin{align}
{\bf E}(x) = [1-w(x)]{\bf E}_{\rm local}(x) + w(x){\bf E}_{\rm eikonal}(x),
\end{align}
where 
\begin{align}
w(x) = \frac{1}{2} \left[1 + \tanh\!\left(2\gamma^{-1/3}( x - x_m)\right)   \right].
\end{align}
The matching point $x_m$ is taken to be a point where both the local and eikonal solutions are valid.  The length scale over which the switching occurs is set using $\gamma^{-1/3}$, since that is a natural length scale in the local solution.  The complex amplitude of the local solution was set so that it matches the eikonal solution at $x=x_m$:
\begin{align}
A_0 = \frac{E_{z,{\rm eikonal}}(x_m)}{\Ai\!\left(-\gamma^{-1/3} (x_m - x_0)\right)}.
\end{align}

\begin{figure}[htb]
\includegraphics[width=8.5cm]{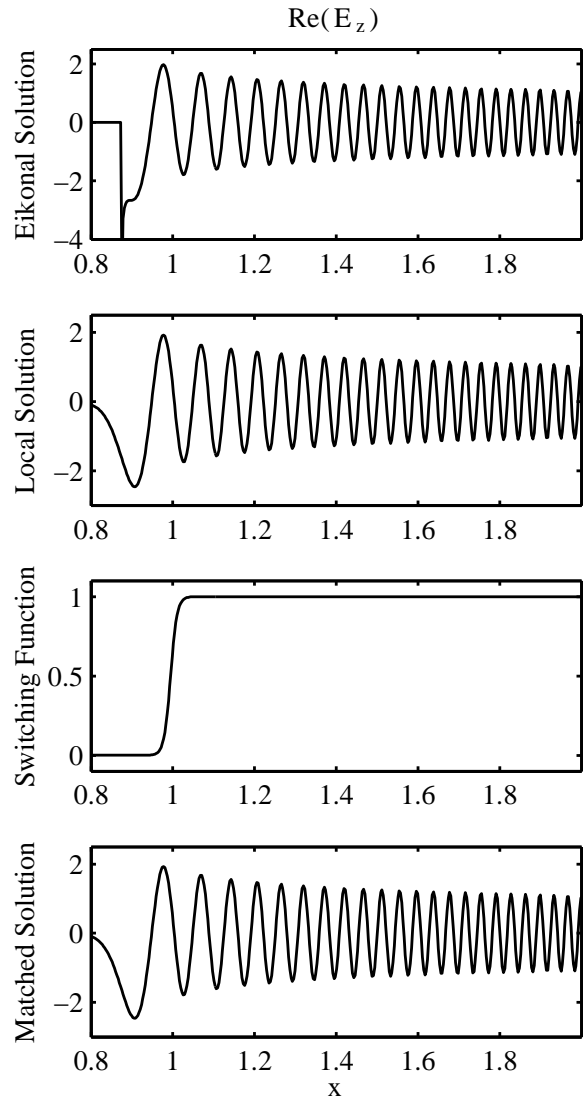}
\caption{The eikonal and local solutions are combined using the switching function to obtain the matched solution.}
\label{fig:1d_eikonal}
\end{figure}

The eikonal field ${\bf E}(x)$ has now been constructed, and gives an approximate field which is good even in the region of the cutoff (see Fig.~\ref{fig:1d_eikonal}).

\subsection{The eikonal solution in two dimensions}

\begin{figure}[tb]
\includegraphics[width=8.5cm]{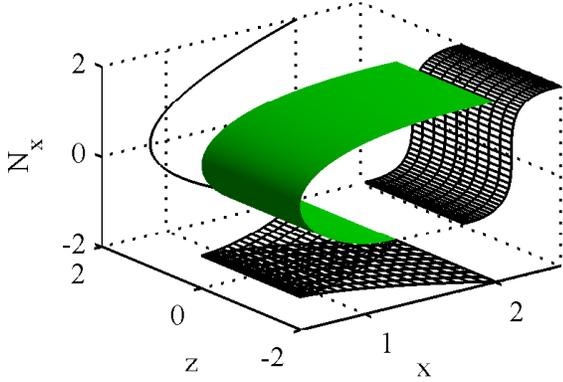}
\caption{(Color online) The family of rays for the 2D case, with projections onto the $(x,z)$-, $(N_x,z)$-, and $(x,N_x)$-planes.}
\label{fig:2d_rays}
\end{figure}

When computing the eikonal field in two dimensions, there are two important aspects which make the calculation more involved than the 1D calculation.  First, instead of simply tracing one ray, and using that to compute the eikonal field, a family of rays must be traced out (Fig.~\ref{fig:2d_rays}).  Information from this family of rays is then used when computing the eikonal quantities $\Theta({\bf x})$, $A({\bf x})$, and ${\bf e}({\bf x})$.  This is for the most a part relatively simple change to implement numerically.  Care must be taken, however, that the initial amplitude and phase at each ray is set appropriately.  The initial phase is set by integrating $({\bf k}\cdot d{\bf x})$ from a chosen reference ray to the initial point of each ray in the family.  The initial amplitude is somewhat more arbitrary.  In this problem, we choose to set the initial amplitude for each ray using a gaussian profile, so that our resulting field will hopefully match the analytical solution shown in Fig.~\ref{fig:exact}.

\begin{figure}[tb]
\includegraphics[width=8.5cm]{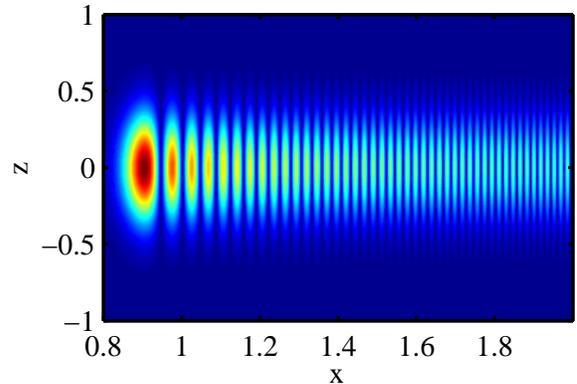}
\caption{(Color online) Absolute value of the $z$ component of the local solution used for matching the eikonal waves in two dimensions.}
\label{fig:2d_local}
\end{figure}

The second critical distinction between the 1D and 2D cases involves the question of how to find the local solution at the caustic.  This is arguably the most difficult aspect of constructing eikonal solutions in two dimensions.  However, in our problem the simple geometry makes the calculation of the local field much easier.  The $x$-dependance of the local solution will be identical to that found for the 1D case.  The $z$-dependance is the simple plane wave $\exp(i k_z z)$ multiplied by the gaussian amplitude profile which comes from the initial conditions (see local solution in Fig.~\ref{fig:2d_local}).

\begin{figure}[tb]
\includegraphics[width=8.5cm]{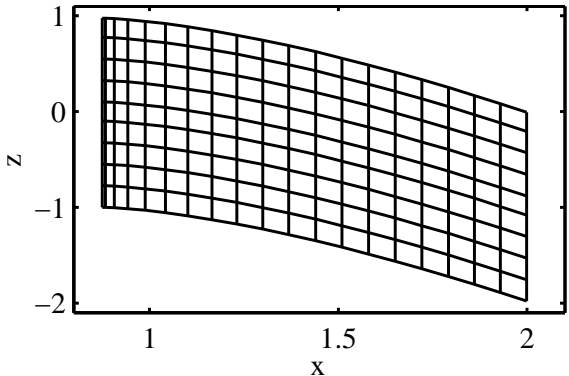}
\caption{Projection of the family of rays onto the $(x,z)$-plane.  The eikonal data is calculated on this mesh.}
\label{fig:2d_grid}
\end{figure}

\begin{figure}[tb]
\includegraphics[width=8.5cm]{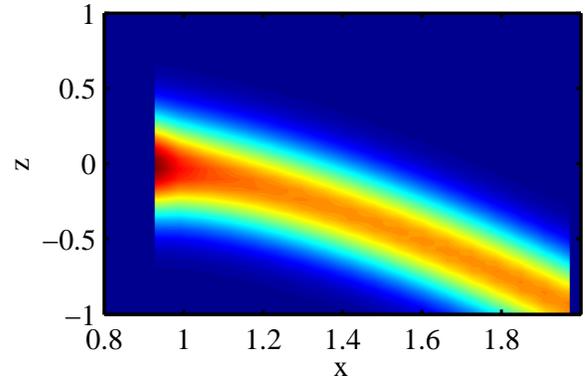}
\caption{(Color online) Eikonal amplitude of the incoming wave, after interpolation onto $(x,z)$.}
\label{fig:2d_amp}
\end{figure}

For this problem, the most computationally expensive part of the 2D calculation was interpolating the amplitude and phase data from functions specified along the rays, to functions of $(x,z)$.  Figure~\ref{fig:2d_grid} shows how the projection of the ray data onto the $(x,z)$-plane forms a nonuniform mesh.  The eikonal data is specified on this mesh, and must be interpolated onto a uniform mesh in $(x,z)$ in order to construct the eikonal solution.  As an example, Fig.~$\ref{fig:2d_amp}$ shows the interpolated amplitude of the incoming eikonal wave.

\begin{figure}[tb]
\includegraphics[width=8.5cm]{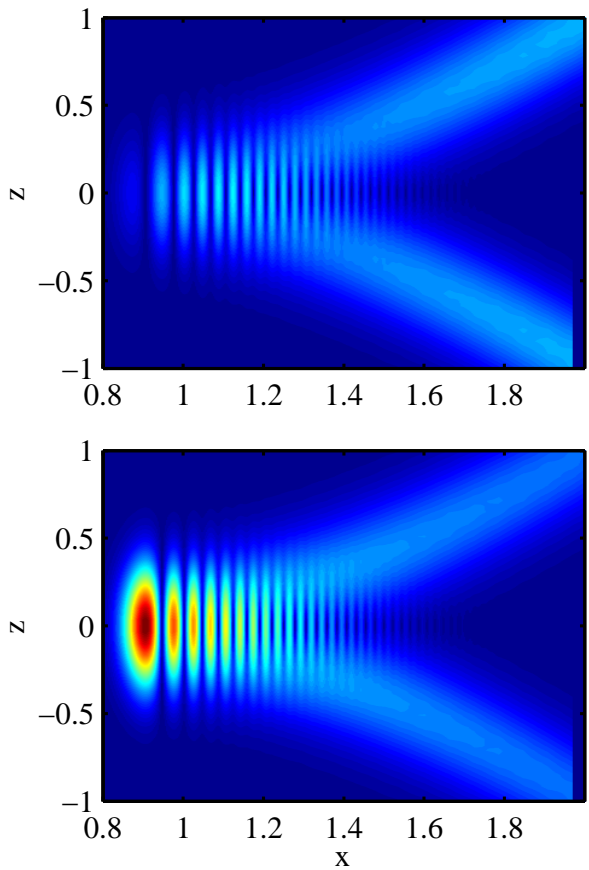}
\caption{(Color online) Absolute value of $E_x$ (top) and $E_z$ (bottom) for the matched eikonal solution.}
\label{fig:2d_eik_sol}
\end{figure}

Once the eikonal phases and amplitudes of the incoming and outgoing waves are interpolated onto the $(x,z)$-plane, we are ready to match them to a local solution, and obtain our final fields.  The same calculation as performed in the previous section gives the phase shift of $\pi/2$ for the outgoing wave.  Matching the eikonal waves to the local solution (Fig.~\ref{fig:2d_local}) which was described above, gives us our final results, shown in Fig.~\ref{fig:2d_eik_sol}.  The result looks quite similar to our analytical solution; a detailed comparison will be made in Section~\ref{sec:compare}.


\section{Semiclassical wave packet dynamics\label{sec:wp}}

In the previous section, the standard eikonal field approximation was computed for the lower hybrid cutoff model.  While the eikonal approximation gives good results for this problem, it is somewhat unwieldy as a numerical algorithm, particularly because of the special care which must be taken in order to get a good field approximation at the cutoff.  The process of finding a local solution, and matching it to incoming and outgoing waves, was fairly straightforward for our slab model, but more complicated geometries would make this process much more difficult.

The standard approach of the previous section is not the only algorithm for constructing eikonal approximations, however.  For example the beam tracing approach described in Ref.~\cite{pereverzev:1998}, has been applied with some success to problems in plasma physics.\cite{maj:062105}  In AMO physics \cite{citeulike:703463,alonso:1699,PhysRevA.40.6814,PhysRevLett.54.1211} and geophysics \cite{thomson_seismic_CS} various methods using the idea of coherent states have been developed.  The advantage of such methods is that they can naturally be interpreted in terms of the phase space dynamics of coherent states, which are a particular type of wave packet.  Because the approximations being used for these methods are based in phase space, they have a natural advantage over configuration-space methods such as the standard eikonal approach described in the last section.  Specifically, the presence of caustics is due to singular projections from phase space to configuration space, and since phase space methods do not require such projections, caustics do not arise in these methods.  While the beam tracing method of Ref.~\cite{pereverzev:1998} can avoid some caustics, it is however still based on constructing a local coordinate frame in configuration space.  This can lead to restrictions on the applicability of the method (e.g., the beam injection angle limit described in Ref.~\cite{maj:062105}) which would not be present for methods based in phase space.

In this section, we will use a technique which was developed as a  semiclassical approximation for quantum problems, and which uses the semiclassical dynamics of wave packets for constructing approximate fields \cite{citeulike:703463}.  This algorithm has a distinct advantage over the standard eikonal field approximation, because it can automatically construct the field in the vicinity of a caustic, with no singularities in the solution.  The amplitude of the solution remains finite at the cutoff, and the phase shift between the incoming and outgoing waves is computed automatically.  This makes the wave packet approach ideally suited for implementing as an eikonal field approximation algorithm.

\subsection{Description of the algorithm\label{sec:wp_algo}}

The algorithm described in this section for field reconstruction was derived for solving quantum mechanical problems using the semiclassical dynamics of wave packets.  As described by Littlejohn in Ref.~\cite{citeulike:703463}, the dynamics of a wave packet can be approximated by considering ray trajectories in phase space.  A wave packet initially centered at the point ${\bf x}_0$ with the carrier wave vector ${\bf k}_0$ will evolve into a wave packet centered at ${\bf x}(t)$ with wave vector ${\bf k}(t)$, where $({\bf x}(t),{\bf k}(t))$ are the coordinates of the ray in phase space.  The width and orientation of the gaussian wave packet then depend on nearby rays.  If the nearby rays are diverging, then the wave packet will spread.  Convergence of rays leads to focusing of the wave packet.  It turns out that information about nearby rays can be approximated by a symplectic matrix ${\bf S}(t)$, and that the dynamical equation for ${\bf S}(t)$ has been derived.\cite{citeulike:703463}

It will be convenient to group the phase space coordinates together as $\xi = ({\bf x},{\bf k})$, with ${\bf x} = (x,z)$ and ${\bf k} = (k_x,k_z)$.  With these definitions, there are three sets of ODEs that need to be solved in order to approximate the dynamics of the wave packet (for the derivation of these equations, see Ref.~\cite{citeulike:703463}):
\begin{eqnarray}
\dot \Theta &=& \frac{1}{2} ( {\bf k} \cdot \dot {\bf x} - {\bf x} \cdot \dot {\bf k}) - D(\xi(t)), \label{eq:ode1}\\
\dot \xi &=& {\bf J} \cdot \nabla D(\xi(t)), \label{eq:ode2}\\
\dot {\bf S} &=& {\bf J} \cdot \nabla \nabla D(\xi(t)) \cdot {\bf S}. \label{eq:ode3}
\end{eqnarray}
In these equations, $\Theta$ is the eikonal phase, $D$ is the dispersion function, and ${\bf J}$ is the symplectic matrix
\begin{align}
{\bf J}=
\left(
\begin{array}{cc}
{\bf 0}  & {\bf id}  \\
-{\bf id}  & {\bf 0}  
\end{array}
\right).
\end{align}
The gaussian wave packet that is following the ray $\xi(t)$ is then given by
\begin{eqnarray}
{\bf E}({\bf x},t) = \frac{ E_0  \hat e(t) e^{i \Theta(t)-\phi(t)}}{\sqrt{\det({\bf A} + i {\bf B})}}  \,\exp\! \big[ i {\bf k}(t) \!\cdot\! \Delta{\bf x}(t) \big], \label{eq:wavepacket}
\end{eqnarray}
where the gaussian envelope of the packet, and the curvature of the phase fronts, are given by
\begin{eqnarray}
\phi(t) =  \frac{1}{2}\Delta{\bf x}(t) \!\cdot\!   ({\bf D}-i{\bf C}) ({\bf A}+i{\bf B})^{-1}  \!\cdot\!  \Delta{\bf x}(t),
\end{eqnarray}
and where
\begin{align}
\Delta{\bf x}(t) = {\bf x}-{\bf x}(t).
\end{align}
The matrices ${\bf A}, {\bf B}, {\bf C}$ and ${\bf D}$ are the submatrices of the symplectic matrix ${\bf S}$:
\begin{eqnarray}
{\bf S} = \left(\begin{array}{cc}
{\bf A} & {\bf B} \\
{\bf C} & {\bf D} \end{array}\right).
\end{eqnarray}
Note that ${\bf x}$ is the point where the field is being evaluated, while ${\bf x}(t)$ is a point on the ray.  The polarization $\hat e (t)$ is the unit zero-eigenvector of the dispersion matrix evaluated at the point $\xi(t)$ on the ray, and $E_0$ is the initial amplitude.  Generically, the matrix ${\bf S}$ (and thus its submatrices ${\bf A}, {\bf B}, {\bf C}$ and ${\bf D}$) must be computed numerically.

There are two important features of Eqn.~(\ref{eq:wavepacket}) for the purpose of implementing it numerically.  First, like other ray tracing algorithms, solving the wave equation has been reduced to solving the ODEs in Eqns.~(\ref{eq:ode1}) and (\ref{eq:ode3}).  [We solve Eqns.~(\ref{eq:ode1}) and (\ref{eq:ode2}) using standard integration algorithms, and Eqn.~(\ref{eq:ode3}) using the algorithm described in App.~\ref{app:a}.]  The second important feature of Eqn.~(\ref{eq:wavepacket}) is the one which makes this method very attractive for numerical implementation.  It has been shown\cite{Littlejohn:1987yf} that the matrix ${\bf A} + i {\bf B}$ is never singular.  This means that its determinant is never zero, and thus the wave packet amplitude never diverges.  This is true even at the location of caustics, where ordinary eikonal methods would break down.  Additionally, the square root $\sqrt{\det({\bf A} + i {\bf B})}$ is a complex function.  By choosing its phase to be continuous in $t$, the phase shift that the wave undergoes at caustics is automatically calculated.  Because Eqn.~(\ref{eq:wavepacket}) is valid and well behaved at caustics, it is ideally suited for implementation as an numerical algorithm for wave field calculations.  Such an algorithm should be able to correctly reconstruct the wave field, even near caustics.  Because of this, it would be superior to the standard eikonal method described in the previous section, where the caustic at the cutoff required special handling.

Fourier transforming the wave packet in Eqn.~(\ref{eq:wavepacket}) from time to frequency gives a time-independent wave field:
\begin{eqnarray}\label{eq:packet_sum}
{\bf E}_\omega({\bf x}) = \int e^{i\omega t} {\bf E}({\bf x},t) \, dt.
\end{eqnarray}
While there is no reason in general that this integral should converge\cite{PhysRevLett.56.2000}, integration over a finite interval in time gives a field which can be interpreted as a ``beam'' of finite length.  In the next section, we show the results of implementing this approximation for the lower hybrid wave reflecting from the cutoff in a plasma.

\subsection{Setting initial conditions}

Perhaps the most difficult issue to address when using the wave packet approximation described in Sec.~\ref{sec:wp_algo} is the question of how to set the initial conditions for the wave packet.  The initial point of the ray sets the initial location of the peak of the wave packet, but the packet shape is set by the initial value of the matrix ${\bf S}$.  Fortunately, this initial value can be set even after ${\bf S}(t)$ is calculated, because of the form of the solution.  Eqn.~(\ref{eq:ode3}) implies that a solution ${\bf S}(t)$ has the form
\begin{eqnarray}
{\bf S}(t) = \exp \!\left\{ {\bf J}\cdot \nabla\nabla D[\xi(t)]  \right\} \cdot {\bf S}_0,
\end{eqnarray}
where ${\bf S}_0 = {\bf S}(0)$ is the initial condition for the matrix ${\bf S}$ \cite{Def:nab2}.  Since the solution has this form, another solution with different initial conditions $\tilde{\bf S}_0$ can be constructed simply by multiplying on the right:
\begin{eqnarray}
\tilde{\bf S}(t) = {\bf S}(t) \cdot {\bf S}_0^{-1} \cdot \tilde{\bf S}_0.
\end{eqnarray}
For our calculations, we used the identity matrix as the initial condition; ${\bf S}_0 = {\bf id}$.  Then, after finding ${\bf S}(t)$, we applied new initial conditions which gave the desired packet shape.

The symplectic matrix ${\bf S}_0$ of initial conditions is a $2n$-dimensional matrix, where $n$ is the number of spatial dimensions.  The space of symplectic matrices has dimension $n(2n+1)=10$.  For a generic problem, the constraints used to set ${\bf S}_0$ would be related to the source of the wave, such as an antenna or a waveguide.  In the problem considered in this paper however, we are studying the reflection of a beam from the cutoff and so the details of the source of the beam are not as important.  Therefore, we are primarily interested in choosing ${\bf S}_0$ so that the resulting beam has a shape which matches our analytical solution.  The choice of ${\bf S}_0$ for a generic problem, and its relation to realistic boundary conditions, is an area for future research.

Numerically, a parameterization such as that described in Ref.~\cite{dopico:650} can used to set ${\bf S}_0$.  However, after exploring various initial conditions, it was found that a relatively simple parameterization could reproduce our analytical solution fairly well: 
\begin{eqnarray}
{\bf S}_0 = 
\left(\begin{array}{cc}
{\bf G} & {\bf 0} \\
{\bf 0} & ({\bf G}^\dagger)^{-1} \end{array}\right),
\end{eqnarray}
where $\bf G$ is an invertible $n\times n$ matrix.  For our problem, we found that a real-valued, diagonal matrix form for $\bf G$ gave us reasonably good results:
\begin{eqnarray}
{\bf G} = 
\left(\begin{array}{cc}
{\sigma_x} & 0 \\
0 & \sigma_z \end{array}\right).
\end{eqnarray}
The parameter $\sigma_z$ was estimated by taking the analytical solution shown in Fig.~\ref{fig:exact}, and fitting a gaussian function of $z$ to the incoming beam.  This process is similar to that which was used to set the initial amplitude for the eikonal solution in the previous section.  Setting the parameter $\sigma_x$ is somewhat more difficult.  Numerically, it was found that varying $\sigma_x$ could significantly change the spreading of the wave packet.  Fairly reasonable results were found by setting $\sigma_x$ in the following way.  

\begin{figure}[tb]
\includegraphics[width=8.5cm]{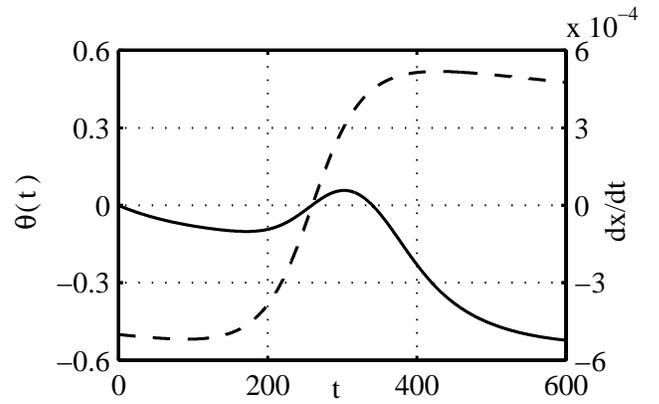}
\caption{The initial value of $\sigma_x$ for the wave packet was set so that the packet hits the cutoff ``head-on,'' i.e., $\theta(t)=0$ (solid line) at the same time that ${dx}/{dt}=0$ (dashed line).}
\label{fig:wp_ini}
\end{figure}

Consider the gaussian envelope of the wave packet at time $t$.  Contours of the wave packet are ellipsoids which make an angle $\theta(t)$ with the $z$-axis.  The value of $\sigma_x$ was chosen so that the wave packet hits the cutoff ``head-on.''  This was done by numerically finding the value of $\sigma_x$ such that $\theta(t^*)=0$, where $t^*$ is the time when the ray reflects from the cutoff.  This value can be found visually by plotting $\theta(t)$ and ${dx}/{dt}$ together.  At the cutoff, ${dx}/{dt}=0$ since the ray reflects in $x$.  So, we varied $\sigma_x$ until $\theta(t)=0$ at the same point that ${dx}/{dt}=0$ (see Fig.~\ref{fig:wp_ini}).  Numerically, this results in the values $\sigma_x=0.1174$ and $\sigma_z=0.2301$.

\subsection{The wave packet solution}

\begin{figure}[tb]
\includegraphics[width=8.5cm]{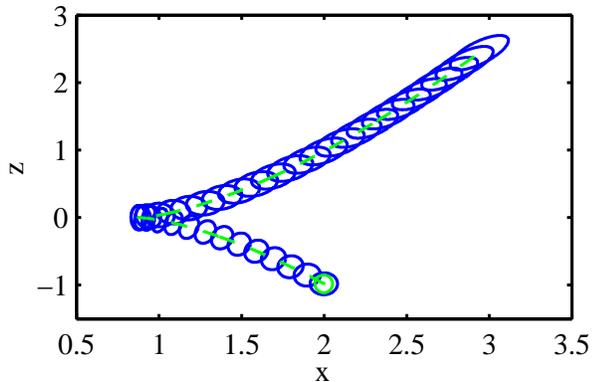}
\caption{(Color online) Ellipses drawn at the 1-$\sigma$ amplitude level show the time dynamics of the wave packet.  Notice the reflection from the cutoff, and the spreading of the wave packet.}
\label{fig:packets}
\end{figure}

\begin{figure}[tb]
\includegraphics[width=8.5cm]{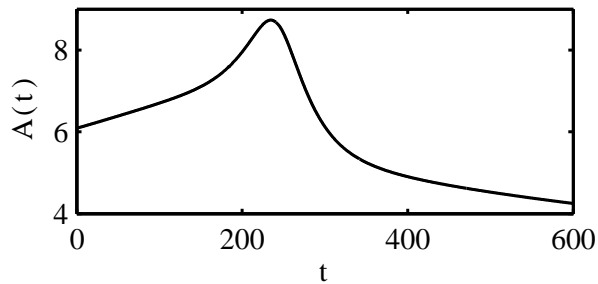}
\caption{Wave packet amplitude, $A= |\det({\bf A} + i {\bf B})|^{-1/2} $ as a function of ray parameter $t$.  The amplitude increases at the cutoff and then decreases as the wave packet spreads.}
\label{fig:amp}
\end{figure}

Having set the initial conditions as described above, the wave packet solution was computed, and the fields were calculated using Eqn.~(\ref{eq:packet_sum}).  Figure \ref{fig:packets} shows 1-$\sigma$ contours of the wave packet's gaussian envelope as a function of time.  Fig.~\ref{fig:amp} shows the wave packet amplitude, $A= |\det({\bf A} + i {\bf B})|^{-1/2} $.  Notice how the amplitude gets large at the cutoff, but does not go to infinity as the eikonal amplitude does.

\section{Comparison of solutions\label{sec:compare}}

In Sec.~\ref{sec:exact}, the analytical solution was found for a lower hybrid wave reflecting from a cutoff in a slab model.  The standard eikonal field approximation was constructed in Sec.~\ref{sec:eikonal}, and in Sec.~\ref{sec:wp}, an alternative algorithm based on the semiclassical dynamics of wave packets was used to obtain an approximate solution.  In this section, we compare the two approximate solutions to the analytical solution, in order to estimate their accuracy.  Note that the eikonal solutions were normalized to match the analytical solution.

\begin{figure}[tb]
\includegraphics[width=8.5cm]{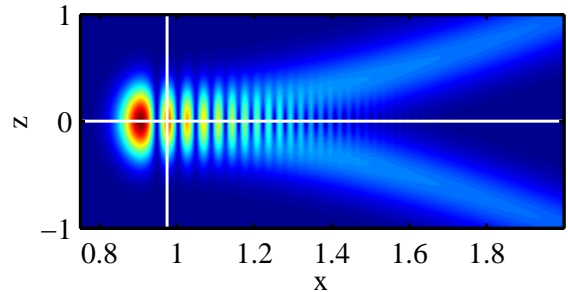}
\caption{(Color online) Locations in the $(x,z)$-plane where the comparisons are performed are marked by white lines.  The $E_z$ component of the field is compared at $x=0.905$, and and at $z=0$.}
\label{fig:comp_loc}
\end{figure}

\begin{figure}[tb]
\includegraphics[width=8.5cm]{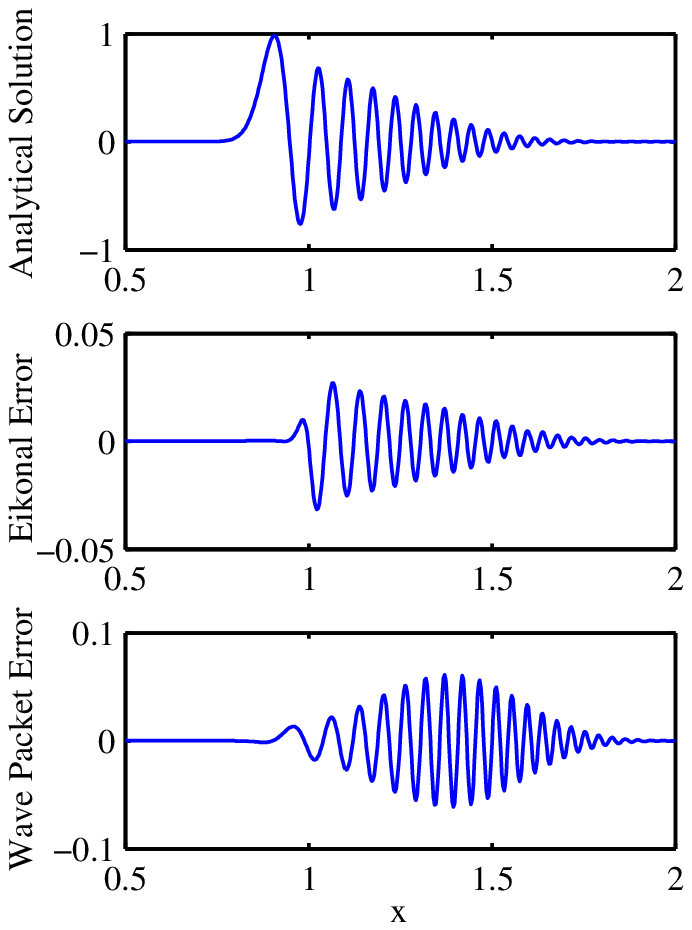}
\caption{Comparison of approximate solutions to the analytical solution.  This is a comparison of the real part of $E_z$, computed at $x=0.905$, showing a slice across the beam.  The top panel shows the exact solution, the middle (bottom) panel shows the error in the eikonal (wave packet) field approximation.}
\label{fig:comparison_x}
\end{figure}

\begin{figure}[tb]
\includegraphics[width=8.5cm]{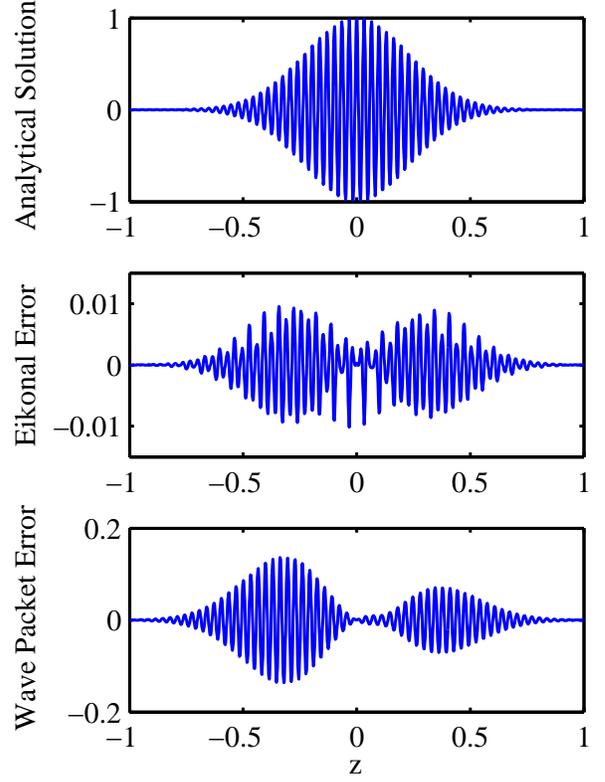}
\caption{Comparison of approximate solutions to the analytical solution.  This is a comparison of the real part of $E_z$, computed at $z=0$.  The top panel shows the exact solution, the middle (bottom) panel shows the error in the eikonal (wave packet) field approximation.}
\label{fig:comparison_z}
\end{figure}

Rather than comparing the entire 2D fields, we compare the fields at fixed $x$ (and $z$) position.  Figure \ref{fig:comp_loc} shows the locations where these comparisons are taken, while the actual comparisons are shown in Figs.~\ref{fig:comparison_x} and \ref{fig:comparison_z}.  As can be seen in the comparisons, the error in the eikonal construction (calculated by subtracting the approximate field from the analytical solution) is about 3\%, while the error in the fields constructed using the wave packet is about 10\%.  Looking at the shape of the error in Fig.~\ref{fig:comparison_z}, it is not unreasonable to suspect that the spreading of the wave packet is causing the error to be this large, since the errors are largest at the edges of the beam.

The spreading seen in the semiclassical wave packet dynamics would be an interesting effect to examine in future research.  The wave packet dynamics uses second derivatives of the dispersion function, and so in a sense incorporates more information about the physics of the problem than the standard eikonal method.  It would be interesting to study how much of the spreading is due to numerical effects, and how much is real physics.  Another aspect of the wave packet method which would be interesting to study would be the setting of the initial condition ${\bf S}_0$ for the wave packet.  Numerical exploration of various values of ${\bf S}_0$ showed that the spreading can change quite a lot depending on the choice of ${\bf S}_0$.  Perhaps a different choice of ${\bf S}_0$ could even reduce the errors seen in Figs.~\ref{fig:comparison_x} and \ref{fig:comparison_z}.  Further study of this aspect of the problem would be needed to determine if this were possible.

\section{Conclusion}

In this paper the reflection of a lower hybrid wave from a cutoff was studied, in order to illustrate two different eikonal methods for constructing approximate solutions.  Comparing these methods in this simple model allowed us to examine the relative merits of each.

The standard eikonal techniques used in Sec.~\ref{sec:eikonal} give very good results, but special care is needed near the cutoff so that the field near the caustic is calculated correctly.  This is relatively straightforward to do in the slab model discussed here, but in more realistic geometry it would be much more difficult.  Also, the calculations needed for calculating the fields near the caustic are difficult to automate on a computer, and thus limit the usefulness of this method for generic eikonal calculations.

The algorithm reported in Sec.~\ref{sec:wp} is based on the semiclassical dynamics of wave packets.  While this algorithm was developed for quantum mechanical calculations, it is also applicable to wave problems in plasma physics.  Because the approximations used in this method are based on consideration of nearby rays in phase space, it does not have the projection singularities at caustics that the standard eikonal method has.  This means that the fields based on this method do not blow up at the cutoff.  More importantly from a numerical viewpoint, it means that no special treatment is needed in order to compute the correct fields at caustics.  Because of this, one would expect that this algorithm would be much better suited for computing eikonal field approximations than the standard method.  There are a few downsides to using this method, however, which would need to be addressed in order to create a robust code based on this algorithm.  In particular, the issue of how to initialize the wave packet in order to get a good match to initial or boundary conditions would have to be addressed.  The method used in this paper turned out to be sufficient for the simple problem considered here, but a better method would need to be developed for more generic problems.  There is also the question of what to do when the spreading of the wave packet becomes too large, causing the nearby ray approximation to break down.  This would be an issue especially in problems where the system size is relatively small, since smaller wave packets tend to spread faster.  It could also be an issue in problems where very long ray trajectories would need to be followed.  Despite these shortcomings, this wave packet algorithm holds promise for performing computations in a variety of realistic plasma physics problems.

The calculations reported here show that it is possible to develop a code to construct approximate wave fields based on the semiclassical dynamics of wave packets.  With care taken to incorporate the effects of antenna geometry, better plasma models (including damping), and more realistic geometry (perhaps even three dimensions), such a code could produce useful and interesting results.

\section*{Acknowledgments}
This research was supported in part by an appointment to the U.S. Department of Energy Fusion Energy Postdoctoral Research Program administered by the Oak Ridge Institute for Science and Education.

\appendix

\section{Solving for ${\bf S}(t)$\label{app:a}}

The symplectic matrix ${\bf S}(t)$ describes the dynamics of nearby rays, and gives the parameters of the gaussian wave packet used for constructing an approximate field in Sec.~\ref{sec:wp}.  The symplectic nature of this matrix is an important property, and it is not preserved by a generic ODE solver.  We used the following algorithm for solving Eqn.~(\ref{eq:ode3}), which has the property that ${\bf S}(t)$ can be kept symplectic to desired order in $\Delta t$.

Rather than numerically solving Eqn.~(\ref{eq:ode3}) for the components of ${\bf S}(t)$, first notice that the solution has an exponential form:
\begin{align}
{\bf S}(t+\Delta t) &= \exp\left( \int_t^{t+\Delta t} {\bf J}\cdot \nabla\nabla D({\bf z}(t')) \, dt' \right) \cdot {\bf S}(t) \\
&\simeq \exp\left({\bf J}\cdot \nabla\nabla D({\bf z}(t)) \Delta t\right) \cdot {\bf S}(t).
\end{align}
This approximation for ${\bf S}$ after a time step $\Delta t$ has a particularly nice property: it can be shown that the matrix $\exp\left({\bf J}\cdot \nabla\nabla D({\bf z}(t)) \Delta t\right)$ is symplectic.  The product of symplectic matrices is also symplectic, so this approximation for ${\bf S}(t+\Delta t)$ is also symplectic.

Now, in order to compute the exponential of a matrix, expand the exponential above to desired order in $\Delta t$:
\begin{align}\label{eq:exp_expand}
{\bf S}(t+\Delta t) \simeq \left( {\bf id} + \Delta{\bf S} + \frac{1}{2!}(\Delta{\bf S})^2 + \ldots \right) \cdot {\bf S}(t),
\end{align}
where
\begin{align}
\Delta{\bf S} = {\bf J}\cdot \nabla\nabla D({\bf z}(t)) \Delta t.
\end{align}
While the expansion in Eqn.~(\ref{eq:exp_expand}) does not maintain the symplectic property which we desired, higher order terms can be included to yield an approximation which is closer to being symplectic.  The error should go like $(\Delta t)^{n+1}$, where $n$ is the highest order kept in the expansion.  For the results reported in this paper, $n=4$.


\begin{thebibliography}{25}
\bibitem{Delos:quote}
\bibinfo{note}{I borrow this phrase from J.~B.~Delos, who uses it to describe the amount of work required to properly use the Maslov methods to reconstruct fields.}
\bibitem{Def:nab2}
\bibinfo{note}{The notation $\nabla\nabla$ in this equation is meant to be interpreted as a tensor valued derivative: $[\nabla\nabla D]_{i,j} = \partial_{\xi_i}\partial_{\xi_j} D$, where $\xi_i$ ($i=1\ldots 2n$) are the $2n$-dimensional phase space coordinates.}
\end{thebibliography}
\end{document}